\documentclass[conference, comsoc]{IEEEtran}
\IEEEoverridecommandlockouts
\usepackage[left=0.625in,right=0.625in,top=0.75in,bottom=1in]{geometry}

\usepackage{caption}
\usepackage{subcaption}
\usepackage{multirow}
\usepackage{array}
\usepackage{balance}
\usepackage{cite}
\usepackage{amsmath,amssymb,amsfonts}
\usepackage{graphicx}
\usepackage{booktabs}
\usepackage{textcomp}
\usepackage{bm}
\usepackage{nicefrac}
\usepackage{mathtools}
\usepackage{comment}
\usepackage{pgfplots}
\usepackage{pgfkeys,pgfmath,pgfcore}
\pgfplotsset{compat=1.18}
\pgfkeys{/pgf/number format/.cd,fixed,precision=2}

\usepackage[ruled, vlined]{algorithm2e}
\usepackage{multirow}
\usepackage{xcolor} 
\usepackage{colortbl}
\usepackage{threeparttable}
\usepackage{dblfloatfix} 

\definecolor{verylightgray}{rgb}{0.9,0.9,0.9}
\usepackage{tikz}
\usetikzlibrary{arrows.meta,bending,calc}
\def\BibTeX{{\rm B\kern-.05em{\sc i\kern-.025em b}\kern-.08em
    T\kern-.1667em\lower.7ex\hbox{E}\kern-.125emX}}

\newif\ifshowtodos
\showtodostrue

\begin{document}

\title{Multimodal Wireless Foundation Models}
\author{\IEEEauthorblockN{ Ahmed Aboulfotouh and
Hatem Abou-Zeid
}
\IEEEauthorblockA{{Department of Electrical and Software Engineering}, 
{University of Calgary}, Canada}
\thanks{
This work was supported by Alberta Innovates, and the Natural Sciences and Engineering Research Council of Canada (NSERC) Discovery Grant RGPIN-2021-04050. }
}

\maketitle

\begin{abstract}

Wireless foundation models (WFMs) have recently demonstrated promising capabilities, jointly performing multiple wireless functions and adapting effectively to new environments.
However, while current WFMs process only one modality, depending on the task and operating conditions, the most informative modality changes and no single modality is best for all tasks. 
WFMs should therefore be designed to accept
multiple modalities to enable a broader and more diverse range
of tasks and scenarios.
In this work, we propose and build the first \emph{multimodal wireless foundation model} capable of processing both raw IQ streams and image-like wireless modalities (e.g., spectrograms and CSI) and performing multiple tasks across both. We introduce masked wireless modeling for the multimodal setting, a self-supervised objective and pretraining
recipe that learns a joint representation from IQ streams
and image-like wireless modalities.
%
We evaluate the model on five tasks across both modality families: image-based (human activity sensing, RF signal classification, 5G NR positioning) and IQ-based (RF device fingerprinting, interference detection/classification).
The multimodal WFM is competitive with single-modality WFMs, and in several cases surpasses their performance. 
Our results demonstrates the strong potential of developing multimodal WFMs that support diverse wireless tasks across different modalities. We believe this provides a concrete step toward both AI-native 6G and the vision of joint sensing, communication, and localization.

\end{abstract}


\section{Introduction}
\label{sec:intro}

Wireless Foundation Models (WFMs) are an emerging paradigm for Wireless AI. They are general AI models, pretrained on diverse wireless data, enabling them to acquire broad and transferable wireless knowledge.
A single WFM can perform multiple functions, and adapt effectively to new scenarios and environments. The multi-task capability improves efficiency, simplifies deployment, and reduces the need for extensive per-task engineering. With these capabilities, WFMs advance the 6G vision of AI-native networks with intelligence embedded from the air interface to the core.

Recent efforts demonstrated WFMs promising capabilities \cite{rfm_gc, aboulfotouh2024building6gradiofoundation, wavesfm, wirelessgpt, contrastivefm, csi2vec, mashaal, wirelessjepa}. 
Early work in this area shows that self-supervised masked modeling can learn general wireless representations which transfer to downstream tasks \cite{rfm_gc, aboulfotouh2024building6gradiofoundation}. 
WavesFM is pretrained with masked modeling to learn general and transferable representations for image-like wireless modalities (e.g., spectrograms, channel-state information (CSI), resource grids). It jointly performs wireless sensing, positioning and communication-related tasks, showing strong out-of-distribution generalization \cite{wavesfm}.
Wireless channel foundation models that operate on CSI and learn general-purpose embeddings for downstream tasks have also been explored in \cite{wirelessgpt, contrastivefm, csi2vec}.
Masked modeling has been used for CSI-based pretraining in \cite{wirelessgpt}. Contrastive learning for CSI modeling has also been applied in \cite{contrastivefm, csi2vec}. CSI2Vec adopts a triplet-based contrastive loss for pretraining and demonstrates downstream transfer to positioning and channel charting tasks \cite{csi2vec}.
For raw in-phase and quadrature (IQ) streams, IQFM is pretrained using a contrastive approach to learn general features for IQ-centric downstream tasks. It performs modulation classification, RF device fingerprinting, and angle-of-arrival estimation. 
It also shows high sample efficiency, adapting to new tasks with few-shot learning \cite{mashaal}. WirelessJEPA demonstrates pretraining on IQ streams using a joint-embedding predictive architecture, and shows generalization across six downstream tasks \cite{wirelessjepa}.

A common limitation of current WFMs is they handle only one modality.
Wireless systems, however, provide various observations (views) of the same over-the-air transmission, obtained at different stages of the physical layer stack, such as raw IQ samples, CSI, bits, and link quality measurements. 
Depending on the task and operating conditions, the most informative modality changes $-$ and no single modality is best for all tasks. WFMs should therefore be designed to accept multiple modalities to enable
a broader and more diverse range of tasks and scenarios. Potential benefits also include improved robustness via redundancy and complementary cues, as well as opportunities for cross-modal self-supervision, which has been shown to improve generalization \cite{crossmodal, multimodalsurvey}. This motivates \emph{\textbf{multimodal wireless foundation models (WFMs)}} that can natively handle multiple input modalities. 
%
%

\begin{figure*}[t!]
    \centering
    \includegraphics[width=0.8\linewidth, keepaspectratio]{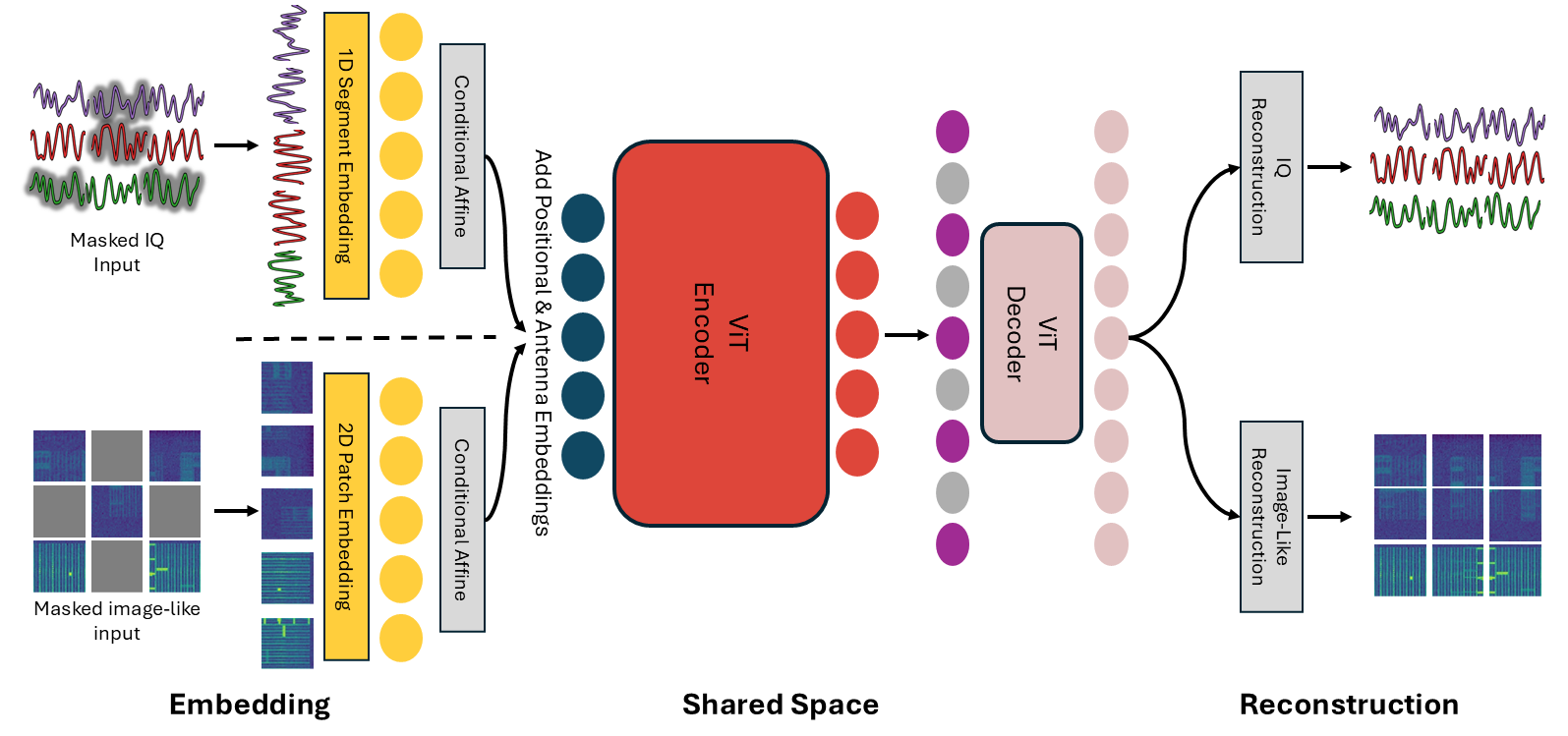}
    \caption{The Proposed Multimodal Wireless Foundation Model.}
    \label{fig:model}
\end{figure*}

In this work, we propose and build the first multimodal WFM that can process raw IQ signals and image-like wireless modalities, and support multiple tasks across both modalities. In more detail, our contributions are:
\begin{itemize}
\item We propose a single \emph{Vision Transformer} (ViT) backbone that supports both modalities via lightweight, modality-specific input embeddings that map inputs into a shared embedding space. This is followed by a shared encoder for feature extraction. The design is masking-friendly and supports multiple task heads.
\item We introduce masked wireless modeling for the \emph{multimodal} setting, a self-supervised objective and pretraining recipe that learns a joint representation from IQ streams and image-like wireless modalities.
\item We provide systematic evidence that masked modeling applied directly to raw IQ streams is effective and yields transferable features that generalize to unseen IQ-centric tasks.
%
\item We evaluate the model under three fine-tuning regimes: linear probing, partial fine-tuning, and low-rank adaptation (LoRA). The evaluation spans five tasks across both modality families: RF fingerprinting, interference detection and classification, human activity sensing, RF signal classification, and 5G NR positioning. The multimodal WFM maintains strong and balanced performance across all tasks, surpassing the IQ baseline WFM on IQ-centric tasks while remaining competitive on tasks with image-like wireless modalities. 
\end{itemize}
\textbf{Reproducible research.} Model weights, finetuning code and reproducibility instructions are available at: https://wavesfm.waveslab.ai.

This work demonstrates the strong capabilities of developing multimodal WFMs that support diverse wireless tasks across different modalities. We believe this provides a concrete step toward both AI-native 6G with the vision of joint sensing, communication, and localization.

\section{Wireless Multimodal Learning Problem}
\label{sec:problem}

\noindent \textbf{Overview.} Consider a wireless network with various nodes: base stations, user equipment, and sensors. These nodes observe propagation events in the radio environment and record various measurements which are used to perform network tasks. Examples of wireless modalities are raw IQ signals, CSI, spectrograms, and link-quality indicators. Nodes consume these modalities to run core functions such as channel estimation, decoding, beamforming, interference mitigation, and link adaptation $-$ in addition to positioning and environment sensing.

\textbf{Problem Definition.} We would like to build a single shared model that can take various modalities and produce a common representation that small neural network task heads can use to solve a range of downstream tasks. The benefit is to cover more functions with one backbone reducing resource demands and keep per-task adaptation minimal.

\textbf{Input families.} We consider two input families: an image-like wireless modality $\mathbf{X}^{\mathrm G}\in\mathbb{R}^{C\times H\times W}$ such as spectrograms and CSI resource grids, where $C$ is the number of channels and $(H, W)$ is the resolution of the image, and a raw IQ stream $\mathbf{X}^{\mathrm Q}\in\mathbb{R}^{M\times T}$ measured over $M$ antennas and $T$ time samples. These measurements arise at different points in the PHY and link stack.

\textbf{Shared model.} A single shared model $f_\theta$ encodes both input families to a modality-agnostic latent representation $\mathbf{z}\in\mathbb{R}^{D_{\text{enc}}}$. Its role is to align features across families while preserving information required by downstream tasks, allowing one backbone to support many functions.

\textbf{Task heads.} For each downstream task $t$ with target space $\mathcal{Y}_t$, a lightweight head $\phi_t:\mathbb{R}^{D_{\text{enc}}}\to\mathcal{Y}_t$ maps the common representation to task outputs $\hat y_t=\phi_t(z)$. These heads are small and supply the minimal adaptation from the shared representation to each task.

\textbf{Learning goal.} The objective is to learn $\theta$ and $\{\phi_t\}$ that achieve strong performance across tasks while keeping the backbone shared and the per-task overhead small:
\begin{equation}
    \label{eq:general-objective}
    \min_{\theta,\{\phi_t\}} \ \sum_{t\in\mathcal{T}} \ \mathbb{E}_{(x,y_t)} \,\ell_t\big(\phi_t(f_\theta(x)),\,y_t\big),
\end{equation}
where $\mathcal T$ is the set of tasks, $x$ is drawn from the input family for task $t$, and $\ell_t$ is the loss for task $t$.

\textbf{Self-supervised learning problem.} Before supervised training of the task heads, we pretrain the shared encoder on unlabeled pools from both families, $\mathcal{D}_{\mathrm G}\subset\mathbb{R}^{C\times H\times W}$ and $\mathcal{D}_{\mathrm Q}\subset\mathbb{R}^{M\times T}$. Pretraining drives the encoder $f_\theta$ to learn robust, transferable features by exploiting structure in unlabeled data, enabling downstream generalization with minimal adaptation and small per-task heads.

\section{Multimodal Wireless Foundation Model}
\label{sec:methods}

In this section, we present our methodology. We first describe the proposed architecture: a ViT-based multimodal design with modality-specific embeddings, a shared encoder for feature extraction, and a lightweight decoder with modality-specific projection heads. We then detail the self-supervised masked wireless modeling objective and pretraining setup, describe the pretraining data, and conclude with the fine-tuning configurations and downstream tasks used for evaluation.

\subsection{Architecture}
We use a masked autoencoder (MAE) with a Vision Transformer (ViT) backbone. It consists of an encoder that maps the masked input to a latent representation and a decoder that reconstructs the original input from this latent. Both the encoder and decoder are shared across modalities. The architecture is illustrated in Fig. \ref{fig:model}.

\textbf{Multimodal Embedding.} Our modalities differ in both dimensionality and statistics, so before the shared encoder we map each one into a common token space. 
For an image-like input $\mathbf{X}^\mathrm{G} \in \mathbb{R}^{C \times H \times W}$, we divide it into a sequence of flattened patches
\begin{equation}
    \label{eq:flattened_patches}
    \mathbf{X}^\mathrm{G}_\mathrm{p} = \left[\mathbf{x}^\mathrm{G}_{\mathrm{p,1}}, \cdots, \mathbf{x}^\mathrm{G}_{\mathrm{p,N}}\right] \in \mathbb{R}^{N\times P^2\cdot C}
\end{equation}
where $N = HW/P^2$ is the number of patches.

For an IQ signal $\mathbf{X}^\mathrm{Q} \in \mathbb{R}^{M \times T}$, we split the complex time series into fixed-length segments of length $S$, and the result is the sequence
\begin{equation}
    \label{eq:segments}
    \mathbf{X}^\mathrm{Q}_\mathrm{s} = \left[\mathbf{x}^\mathrm{Q}_{\mathrm{s,1}}, \cdots, \mathbf{x}^\mathrm{Q}_{\mathrm{s,K}}\right] \in \mathbb{R}^{K \times S}
\end{equation}
where $K = MT/S$ is the number of resulting segments.
We then apply a separate linear projection to each modality to obtain tokens in the encoder dimension $D_{\text{enc}}$.

After projection, each token passes through a modality-specific feature-wise affine layer (learnable scale and shift) as follows:
\begin{align}
    \label{eq:aff_grid}
    \mathbf{U}^\mathrm{G} &= \mathbf{X}^\mathrm{G}_\mathrm{p} \mathbf{E}^\mathrm{G} \odot \mathbf{1}_{N} \big(\bm{\gamma}_\mathrm{G}\big)^\top+ \mathbf{1}_{N} \big(\bm{\beta}_\mathrm{G}\big)^\top \in \mathbb{R}^{N \times D_{\text{enc}}} \\
    \label{eq:aff_iq}
    \mathbf{U}^\mathrm{Q} &= \mathbf{X}^\mathrm{Q}_\mathrm{s} \mathbf{E}^\mathrm{Q} \odot \mathbf{1}_{K} \big(\bm{\gamma}_\mathrm{Q}\big)^\top + \mathbf{1}_{K} \big(\bm{\beta}_\mathrm{Q}\big)^\top \in \mathbb{R}^{K \times D_{\text{enc}}}
\end{align}
where $\mathbf{E}^\mathrm{G}\in\mathbb{R}^{P^2\cdot C \times D_{\text{enc}}}$ and $\mathbf{E}^\mathrm{Q}\in\mathbb{R}^{S \times D_{\text{enc}}}$ are linear projection matrices, $\mathbf{1}_{N}$ denotes the length $N$ column vector of ones, $\bm{\gamma}_\mathrm{G}, \bm{\gamma}_\mathrm{Q} \in \mathbb{R}^D_{\text{enc}}$ are scaling parameters and $\bm{\beta}_\mathrm{G}, \bm{\beta}_\mathrm{Q} \in \mathbb{R}^D_{\text{enc}}$ are shift parameters for IQ and image-like modalities, respectively.
This conditioning step with the separate affine transforms aligns tokens from different modalities into a common representation space, while preserving their unique characteristics.

\begin{algorithm}[t]
\DontPrintSemicolon
\caption{Multimodal Masked Wireless Modeling}
\label{alg:pretraining}
\SetKwInOut{Input}{Input}\SetKwInOut{Output}{Output}\SetKwInOut{Initialization}{Initialization}
\Input{Image-like and IQ datasets $\mathcal{D}_{\mathrm G},\ \mathcal{D}_{\mathrm Q}$, MAE $\mathcal{M}$, mask ratios $\delta_{\mathrm G},\delta_{\mathrm Q}$}
\Output{foundation model $\mathcal{F}$}
\Initialization{$\mathcal{F} \gets $ ViT encoder from $\mathcal{M}$, $P \gets$ patch size, $S \gets$ segment size.}
\Repeat{\normalfont convergence is reached or another stopping condition is met}{
\nl{}\For{\normalfont paired mini-batches $\big(\mathcal{B}_{\mathrm G},\mathcal{B}_{\mathrm Q}\big)$ in $\big(\mathcal{D}_{\mathrm G},\mathcal{D}_{\mathrm Q}\big)$}{
\nl{}Initialize the loss value $\mathcal{L}\gets 0$\\
    \nl{}\For{\normalfont each $\big(\mathbf{X}^{\mathrm G}, \mathbf{X}^{\mathrm Q}\big)$ in $\big(\mathcal{B}_{\mathrm G}, \mathcal{B}_{\mathrm Q}\big)$}{
    \nl{}Map inputs to a sequence of $P\times P$ flattened patches and length $S$ segments $\mathbf{X}^\mathrm{G}_\mathrm{p},\ \mathbf{X}^\mathrm{Q}_\mathrm{s}$ as in equations \eqref{eq:flattened_patches} and \eqref{eq:segments}. \\
    \nl{}Perform the embedding for each sequence to get $\mathbf{U}^{\mathrm G}, \mathbf{U}^{\mathrm Q}$ as in equations \eqref{eq:aff_grid} and \eqref{eq:aff_iq}.\\
    \nl{}Add positional and antenna embeddings to get $\mathbf{Z}^\mathrm{G}, \mathbf{Z}^\mathrm{Q}$ as in equations \eqref{eq:pos_grid}, \eqref{eq:pos_iq}. \\
    \nl{}Mask the token sequences to get $\tilde{\mathbf{Z}}^{\mathrm G}, \tilde{\mathbf{Z}}^{\mathrm Q}$ as in equations \eqref{eq:mask_grid} and \eqref{eq:mask_iq}. \\
    \nl{}Compute the model output for each modality to get the reconstructed patches/segments. \\
    \Indp
    $\hat{\mathbf{X}}^\mathrm{G}_\mathrm{p} \gets \mathcal{M}\big(\mathbf{Z}^\mathrm{G}\big)$; $\hat{\mathbf{X}}^\mathrm{Q}_\mathrm{s} \gets \mathcal{M}\big(\mathbf{Z}^\mathrm{Q}\big)$ \\
    \Indm
    \nl{}Accumulate the per-modality losses $\mathcal{L}_\mathrm{G},\ \mathcal{L}_\mathrm{Q}$ as in equations \eqref{eq:recon_grid} and \eqref{eq:recon_iq}.\\
    \Indp
    $\mathcal{L} \gets \mathcal{L} + \mathcal{L}_ \mathrm G + \mathcal{L}_ \mathrm Q$
    \Indm 
    }
    \nl{}Update parameters of the model $\mathcal{M}$ using the average loss $\mathcal{L}/\big(|\mathcal{B}_{\mathrm G}|+|\mathcal{B}_{\mathrm Q}|\big)$ via backpropagation. \\
}
}
\end{algorithm}

\textbf{Positional \& Antenna Embeddings.}
We add sinusoidal positional embeddings, 2D for image-like inputs (spatial layout) and 1D for IQ (temporal order).
For IQ, we further introduce \emph{antenna embeddings}: in multi-antenna systems, signals at different antennas are related through array geometry, propagation environment, and any beamforming in use. These relationships cannot be captured by sinusoidal embeddings alone. 
To make antenna identity explicit, we add a learned per-antenna vector to every IQ token from its source antenna; the same vector is reused across time and learned jointly with the model.

Let $\mathbf{e}_1,\ldots,\mathbf{e}_M\in\mathbb{R}^D_{\text{enc}}$ be the antenna embeddings and let $a_k\in\{1,\ldots,M\}$ index the source antenna of the $k$-th IQ token.
The resulting tokens are
\begin{align}
    \label{eq:pos_grid}
    \mathbf{Z}^{\mathrm{G}} &= \mathbf{U}^{\mathrm{G}} + \mathbf{E}_{\text{pos}}^{\mathrm{G}} \in \mathbb{R}^{N\times D_{\text{enc}}} \\
    \label{eq:pos_iq}
    \mathbf{Z}^{\mathrm{Q}} &= \mathbf{U}^{\mathrm{Q}} + \mathbf{E}_{\text{pos}}^{\mathrm{Q}} + \mathbf{E}_{\text{ant}} \in \mathbb{R}^{K\times D_{\text{enc}}}
\end{align}
where $\mathbf{E}_{\text{pos}}^{\mathrm{G}}\in\mathbb{R}^{N\times D_{\text{enc}}}$ and $\mathbf{E}_{\text{pos}}^{\mathrm{Q}}\in\mathbb{R}^{K\times D_{\text{enc}}}$ are 2D and 1D sinusoidal positional embeddings, respectively, and
$(\mathbf{E}_{\text{ant}})_{k} = \mathbf{e}_{a_k}^{\top}.$ For details on sinusoidal positional embeddings, see \cite{vit, wavesfm}.

\textbf{Masking.} For each modality, we mask a fraction of the token sequence with $\delta_\mathrm{G}$, $\delta_\mathrm{Q}$ being the mask ratios for image-like and IQ inputs, respectively. The tokens to be removed, are sampled randomly from each sequence. The resulting sequences are
\begin{align}
    \label{eq:mask_grid}
    \tilde{\mathbf{Z}}^{\mathrm G} = \big[\tilde{\mathbf{z}}_{1}^{\mathrm G}, \cdots, \tilde{\mathbf{z}}_{\tilde{N}}^{\mathrm G} \big] \in \mathbb{R}^{\tilde{N} \times D_{\text{enc}}} \\
    \label{eq:mask_iq}
    \tilde{\mathbf{Z}}^{\mathrm Q} = \big[\tilde{\mathbf{z}}_{1}^{\mathrm Q}, \cdots, \tilde{\mathbf{z}}_{\tilde K}^{\mathrm Q} \big] \in \mathbb{R}^{\tilde{K} \times D_{\text{enc}}}
\end{align}
where $\tilde N = N - \lfloor \delta_\mathrm G N\rfloor$ and $\tilde K = K - \lfloor \delta_\mathrm Q K\rfloor$ are the number of remaining tokens, for the image-like and IQ input, respectively.

\textbf{Autoencoder Design.} The encoder operates on the set of visible (unmasked) tokens, processing them through a series of ViT blocks to output feature tokens. These tokens are then projected to the decoder dimension, after which mask tokens are inserted at the previously masked positions. The mask token is a single learnable embedding shared across all masked positions, indicating where reconstruction is required. Positional embeddings are then added to the sequence to preserve ordering, which is then processed by the decoder through a series of ViT blocks. 
The design is asymmetric: most of the computation budget is in the encoder. This allows using the decoder for reconstruction during pretraining only and it can be safely discarded afterward. We do not show the details of ViT forward pass. For more details, refer to \cite{vit, wavesfm}.

\textbf{Reconstruction.}
The decoder outputs are mapped back to the signal domain by separate linear projection heads which are not shared across modalities. Each head converts decoder tokens to the original dimensionality and shape of its modality. Assume the decoder outputs $\mathbf{Y}^{\mathrm G} \in \mathbb{R}^{N \times D_{\text{dec}}}, \mathbf{Y}^{\mathrm Q} \in \mathbb{R}^{K \times D_{\text{dec}}}$ for each modality where $D_{\text{dec}}$ is the decoder dimension, the operation of the per-modality heads can be written as
\begin{align}
    \label{eq:recon_grid}
    \hat{\mathbf{X}}^{\mathrm G}_{\mathrm p} &= \mathbf{Y}^{\mathrm G} \mathbf{E}^{\mathrm G} \in \mathbb{R} ^ {N \times P^2 \cdot C}\ \\
    \label{eq:recon_iq}
    \hat{\mathbf{X}}^{\mathrm Q}_{\mathrm s} &= \mathbf{Y}^{\mathrm Q} \mathbf{E}^{\mathrm Q} \in \mathbb{R}^{K \times S}.
\end{align}
Here, $\mathbf{E}^{\mathrm G} \in \mathbb{R}^{D_{\text{dec}} \times P^2 \cdot C}$ and $\mathbf{E}^{\mathrm Q} \in \mathbb{R}^{D_{\text{dec}} \times S}$ are the projection matrices; $\hat{\mathbf{X}}^{\mathrm G}_{\mathrm p}=\big[\hat{\mathbf{x}}^\mathrm{G}_{\mathrm{p},1}, \ldots, \hat{\mathbf{x}}^\mathrm{G}_{\mathrm{p},N}\big]$ where $\hat{\mathbf{x}}^\mathrm{G}_{\mathrm{p},n} \in \mathbb{R}^{P^2 \cdot C}$ is a single reconstructed patch; and $\hat{\mathbf{X}}^{\mathrm Q}_{\mathrm s}=\big[\hat{\mathbf{x}}^\mathrm{Q}_{\mathrm{s},1}, \ldots, \hat{\mathbf{x}}^\mathrm{Q}_{\mathrm{s},K}\big]$ where $\hat{\mathbf{x}}^\mathrm{Q}_{\mathrm{s},k} \in \mathbb{R}^{S}$ is a single reconstructed segment.

%
Our loss function is the mean squared error (MSE) between the reconstructed and original input, computed only on the masked parts. Let $\mu_\mathrm G \subset\{1, \cdots, N\}$, $\mu_\mathrm Q \subset\{1, \cdots, K\}$ be the indices of the masked tokens of the image-like and IQ inputs, respectively. The per-modality losses $\mathcal{L}_\mathrm G,\ \mathcal{L}_\mathrm Q$ are
\begin{align}
    \label{eq:loss_images} 
    \mathcal{L}_\mathrm G  = \dfrac{1}{|\mu_\mathrm G|} \sum_{i \in \mu_\mathrm G}\big\|\mathbf{x}^\mathrm{G}_{\mathrm{p,i}} - \hat{\mathbf{x}}^\mathrm{G}_{\mathrm{p,i}} \big\| ^ 2\ \\
    \label{eq:loss_iq}
    \mathcal{L}_\mathrm Q  = \dfrac{1}{|\mu_\mathrm Q|}\sum_{i \in \mu_\mathrm Q}\big\|\mathbf{x}^\mathrm{Q}_{\mathrm{s,i}} - \hat{\mathbf{x}}^\mathrm{Q}_{\mathrm{s,i}} \big\| ^ 2.
\end{align}
The end-to-end pretraining is detailed in Algorithm \ref{alg:pretraining}.

\subsection{Pretraining Data}

\begin{figure}[t]
    \centering
    \begin{subfigure}[h!]{\linewidth}
        \centering
        \includegraphics[width=0.85\linewidth, keepaspectratio]{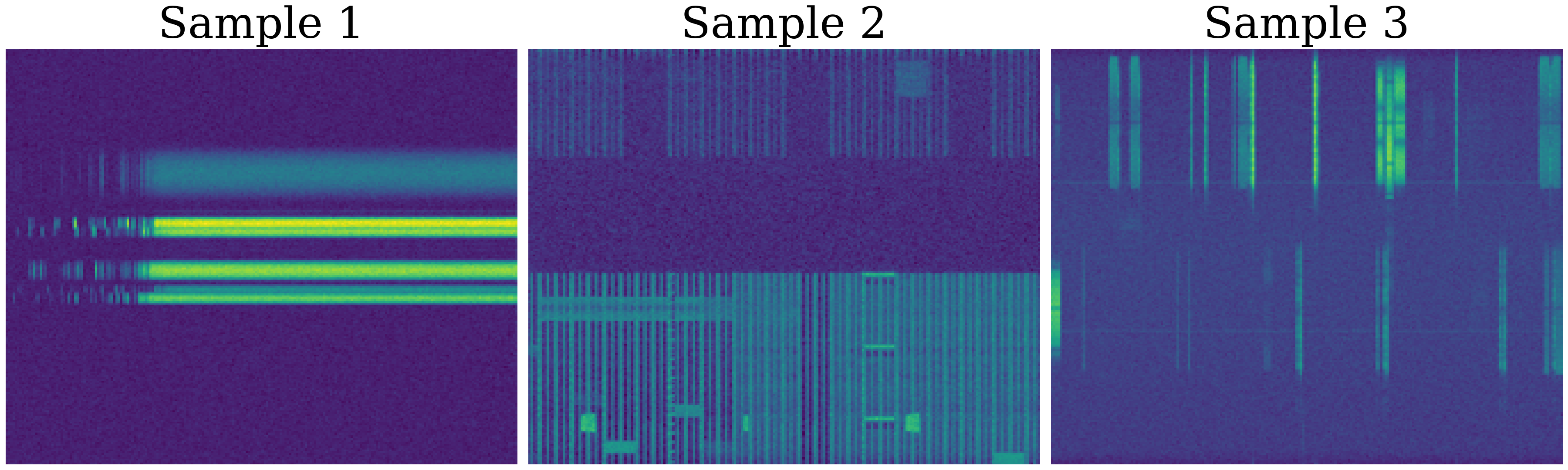}
        \caption{Samples from the spectrogram dataset.}
        \label{fig:spec-data}
        \vspace{2pt}
    \end{subfigure}
    \begin{subfigure}[h!]{\linewidth}
        \centering
        \includegraphics[width=0.85\linewidth, keepaspectratio]{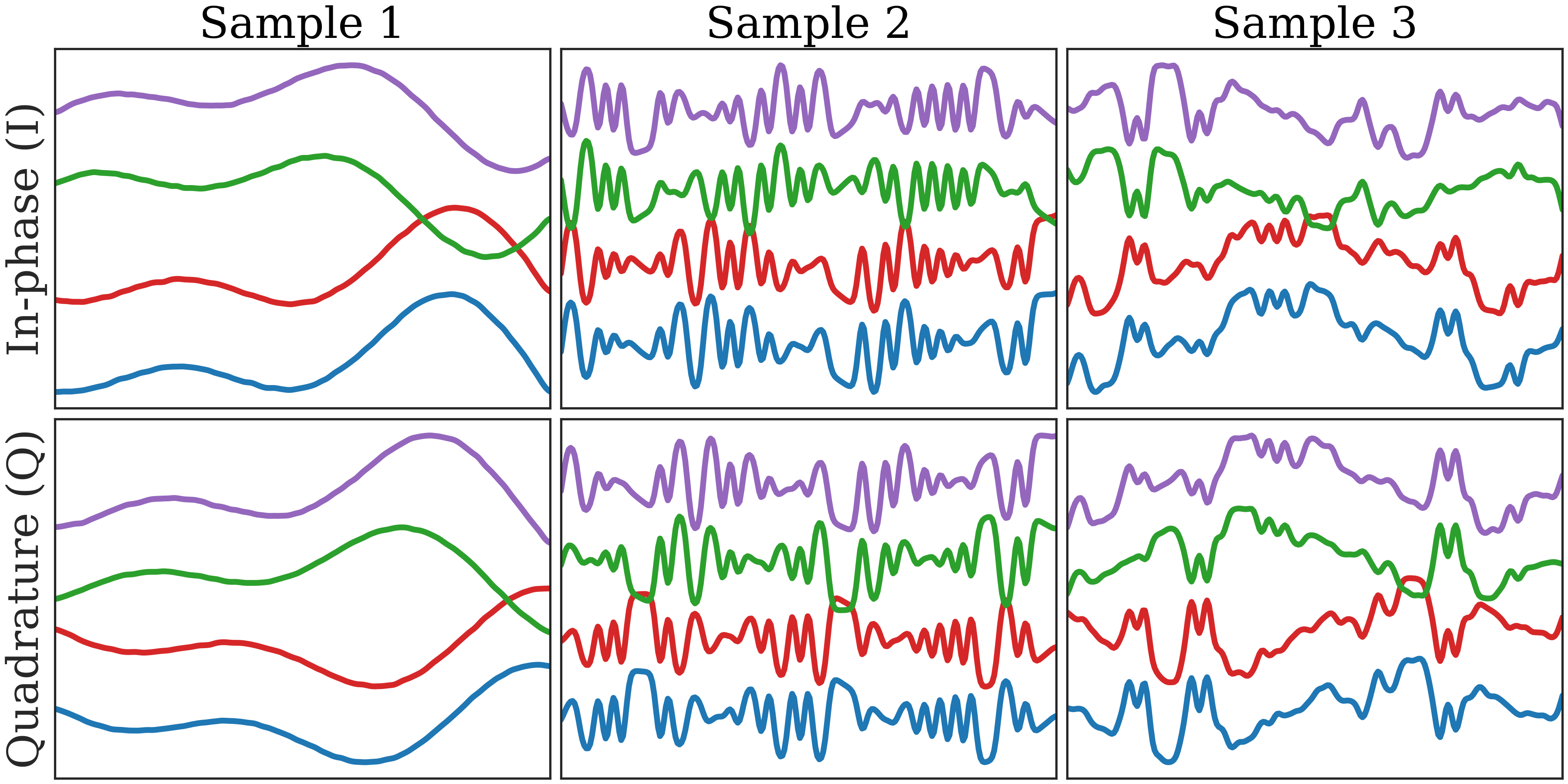}
        \caption{Samples from the IQ dataset. Each sample contains four streams (one per antenna) shown stacked vertically.}
        \label{fig:iq-data}
    \end{subfigure}
    \caption{Pretraining Datasets.}
    \label{fig:datasets}
\end{figure}

We pretrain on two datasets: a spectrogram dataset for image-like wireless modalities and an IQ dataset for raw in-phase and quadrature streams. Samples from the datasets are shown in Fig. \ref{fig:datasets}. In the following, we describe the details of each dataset.

\textbf{Spectrogram Dataset.} The dataset represents image-like wireless modalities. It is a collection of spectrograms for various signal types and bands (e.g., WiFi, LTE, Bluetooth, 5G-NR, and ISM). Recordings are captured over the air using SDRs at various center frequencies in the sub-6 GHz band, and sampling frequency in the range $\big[10,\ 60\big]$ MHz and converted to spectrogram representations; see \cite{rfm_gc} for details of spectrogram computation. The dataset contains a total of 3200 samples.

\textbf{IQ Dataset.} The datasets represents the IQ modality. It consists of baseband IQ recordings from a 4-antenna MIMO setup in an indoor testbed, where signals with different modulations and technologies are exchanged between a transmitter and a receiver under various antenna configurations and transmitter/receiver locations; see \cite{mashaal} for more details about the setup. The dataset contains a total of 3200 samples.

\textbf{Data Preprocessing.} For spectrogram inputs, the steps are: (1) map the spectrogram to the logarithmic scale to reduce sparsity, (2) normalize to the range $[0, 1]$ using dataset-wide statistics, (3) resize to the shape $224 \times 224$, and (4) standardize using dataset-wide statistics. For IQ inputs, we standardize the in-phase and quadrature components using dataset-wide statistics.

\subsection{Fine-Tuning Configurations}

Across all tasks we attach a single linear head sized to the task’s label space. The MAE decoder is used only during pretraining and is not part of inference. Unless stated otherwise, the modality-specific input projections (patch/segment) and antenna embeddings remain \emph{trainable in all regimes}, while positional embeddings are completely frozen.

\textbf{Linear probe (LP).}
The ViT encoder is frozen end-to-end; only the task head and the input modality projections are updated. This setting evaluates the transferability of the pretrained features with minimal adaptation.

\textbf{Partial fine-tuning ($k$ blocks).}
The last $k$ ViT blocks (attention, MLP, and their LayerNorms) are unfrozen; earlier blocks remain frozen. The task head and the input modality projections are trainable. This enables limited adaptation while preserving most pretrained structure.

\textbf{Low-Rank Adaptation (LoRA) adapters.}
Low-rank adapters are inserted into the query and value projections of each encoder block. All original encoder weights remain frozen; trainable parameters are the LoRA adapters, the task head, and the input modality projections. This enables efficient adaptation while keeping the encoder shared among tasks.


\subsection{Downstream Tasks}

We evaluate the model performance on five downstream tasks: two with IQ signals 
and three with image-like wireless modalities.
For classification problems, our optimization objective is the cross-entropy loss, while for regression, we utilize the mean-squared error. 

\textbf{RF Fingerprinting (RFP).} 
Raw IQ traces of LTE and 5G-NR transmissions collected at a fixed receiver from four different base stations. The target is device identification, i.e., which transmitter generated each segment \cite{rfp}.

\textbf{Interference Detection (INTD) and Classification (INTC).}
Over-the-air LTE IQ recordings under clean conditions and under direct-sequence spread-spectrum (DSSS) interference. The task is to detect the presence of interference and, when present, identify its modulation type directly from IQ \cite{icarus}.

\textbf{Human Activity Sensing (HAS).}
Wi-Fi CSI measurements collected while human subjects perform six activities between two access points with three antennas each. Input to the model is the CSI observation and the target is activity label \cite{csi_sensing}.

\textbf{RF Signal Classification (RFS).}
Spectrogram measurements for various signal types, such as WiFi, FM radio, cellular, Bluetooth, and others, with a total of 20 classes. The task is to predict the signal type from its spectrogram \cite{Zahid2024}.

\textbf{5G NR Positioning (POS).}
CSI measurements collected using sounding reference signals in a 5G NR setup. The target is to predict UE location from the CSI observations \cite{5g_positioning}.

\section{Results and Discussion}
\label{sec:results}

In this section, we first outline the pretraining setup and model hyperparameters. We then benchmark the multimodal WFM on downstream tasks against single-modality WFM baselines (WavesFM and IQFM). Finally, we assess its maximal multi-task performance using LoRA adaptation.

\subsection{Experimental Configuration}

\begin{figure}[h!]
    \centering
    \includegraphics[width=0.93\linewidth]{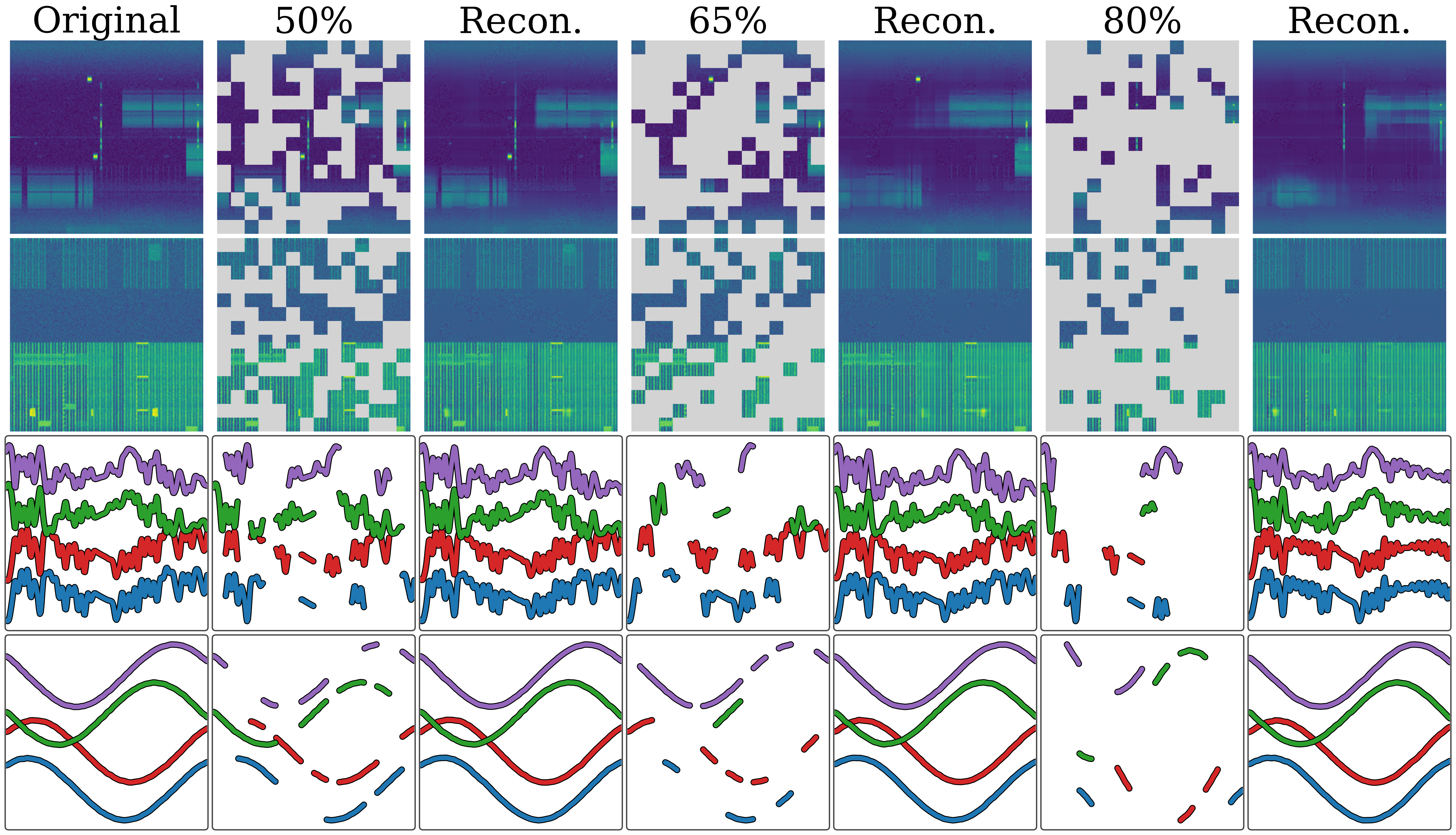}
    \caption{Reconstruction examples at different masking ratios.}
    \label{fig:recon}
\end{figure}

\textbf{Training Setup.}
Pretraining runs for 800 epochs with a 40-epoch linear warm-up. We use a fixed 70\% masking ratio for both modalities. Optimization uses Adam (learning rate \(1\times10^{-3}\)) with cosine annealing. Sampling alternates between modalities in a strict round-robin schedule; one optimization step is taken only after processing a mini-batch from \emph{each} modality, and the two per-modality gradients are aggregated to form the update as in Algorithm \ref{alg:pretraining}. The model configuration is summarized in Table \ref{tab:model-hparams}. We show reconstruction examples by the pretrained model for both modalities in Fig. \ref{fig:recon}.

\begin{table}[h!]
\centering
\caption{Model Hyper-parameters}
\label{tab:model-hparams}
\setlength{\tabcolsep}{8pt}
\renewcommand{\arraystretch}{1.15}
\begin{tabular}{lcc}
\toprule
\textbf{Hyperparameter} & \textbf{ViT Encoder} & \textbf{ViT Decoder} \\
\midrule
Image Patch Size        & 16 $\times$ 16 & -- \\
IQ Segment Size         & 16 & -- \\
Blocks                  & 8  & 4  \\
Embed Dim.              & 256 & 128\\
Hidden Dim.             & 1024 & 512 \\
Attention Heads         & 8  & 16 \\
Parameters (M)          & 6.32 & 0.79 \\
\end{tabular}
\end{table}

\subsection{Baseline Downstream Performance}

\begin{table*}[t!]
\centering
\caption{Performance of the Multimodal WFM compared to baselines on downstream tasks.}
\label{tab:avg-metrics}
\small
\setlength{\tabcolsep}{12pt}
\renewcommand{\arraystretch}{1.15}

\begin{subtable}{\textwidth}
\centering
\caption{Linear probing (LP)}
\label{tab:avg-metrics-lp}
\begin{tabular}{l|ccc|ccc}
\toprule
\textbf{Model} & \textbf{RFP} & \textbf{INTD} & \textbf{INTC} & \textbf{HAS} & \textbf{RFS} & \textbf{POS} \\
\midrule
Multimodal WFM & 98.65 & 96.40 & 58.25  & 95.58 & 63.59  & 2.11  \\
IQFM/WavesFM & 91.41 & 95.00 & 61.72 & 97.7 & 68.10 & 3.90 
\end{tabular}
\end{subtable}

\vspace{0.6em}

\begin{subtable}{\textwidth}
\centering
\caption{Partial fine-tuning of two blocks (FT2)}
\label{tab:avg-metrics-ft2}
\begin{tabular}{l|ccc|ccc}
\toprule
\textbf{Model} & \textbf{RFP} & \textbf{INTD} & \textbf{INTC} & \textbf{HAS} & \textbf{RFS} & \textbf{POS} \\
\midrule
Multimodal WFM & 99.69 & 98.74 & 64.15 & 98.73 & 84.14 & 1.27 \\
IQFM/WavesFM & 96.47 & 96.40 & 62.24 & 98.86 & 86.05 & 0.41 \\
\end{tabular}
\end{subtable}
\end{table*}

We evaluate the performance of the multimodal WFM compared to baselines across the five downstream tasks. Our baselines are modality-specific WFMs: WavesFM (baseline for image-like modalities) and IQFM (baseline for IQ streams). We report results on two finetuning regimes: linear probing (LP) and partial finetuning with 2 blocks (FT2). LP assesses the quality of the learned representation with almost no task specific adaptation, while FT2 is a more flexible setting that allows more adaptation. Results are shown in Table \ref{tab:avg-metrics}. We report mean per-class accuracy 
for classification tasks, and mean localization error for the positioning task. 

On IQ-centric tasks, the multimodal WFM consistently outperforms IQFM under both LP and FT2, with the exception of INTC under LP, showcasing strong generalization.
Because IQFM is a compact model ($\sim0.3$M parameters) compared to the multimodal WFM ($\sim7$M), model capacity should be considered when reading the gaps. It is also reported in \cite{mashaal} that IQFM significantly improves when fine-tuned with LoRA than with LP or partial finetuning.

On image-like tasks, WavesFM leads on RFS, while HAS is close, and the multimodal WFM leads on POS. Under FT2, the RFS gap narrows and WavesFM gains more on POS. Overall, WavesFM is stronger with FT2, but the margins are reasonable. These gaps may reflect differences in pretraining data size and diversity: WavesFM is pretrained on spectrograms and CSI, whereas the multimodal WFM is pretrained on spectrograms only, when it comes to image-like modalities. WavesFM is also about five times larger in parameter count. These factors contribute to the gaps rather than inherent limits of the approach.

Overall, the multimodal WFM delivers strong, balanced performance across \emph{both} modalities: it consistently surpasses IQFM and remains competitive with WavesFM. These findings illustrate the potential of the approach, and scaling up the pretraining corpus (e.g., incorporating CSI) should further improve performance and generalization.

\subsection{Multi-task Performance with LoRA}

LoRA adapts the network by inserting low-rank modules, allowing both early and late blocks to change behavior while the backbone remains frozen. By placing these modules in attention and MLP layers throughout the encoder stack, both shallow and deep features can be adjusted.
Compared to FT2, this is more flexible, keeps the encoder fully shared across tasks, and requires far fewer task-specific parameters. Our prior work showed that LoRA can exceed partial finetuning at a fraction of the parameter cost \cite{wavesfm}. 
We fine-tune with LoRA rank $R=32$ and scaling $\alpha=32$, totaling $\sim\!0.3$M task-specific parameters, versus $\sim\!1.65$M for FT2. Results are reported in Table \ref{tab:wfm-lora}.
\begin{table}[t]
\centering
\caption{LoRA ($R=32,\ \alpha=32$) performance of the multimodal WFM. }
\label{tab:wfm-lora}
\setlength{\tabcolsep}{8pt}
\renewcommand{\arraystretch}{1.15}
\begin{tabular}{ccc|ccc}
\toprule
\textbf{RFP} & \textbf{INTD} & \textbf{INTC} & \textbf{HAS} & \textbf{RFS} & \textbf{POS} \\
\midrule
99.77 $\uparrow$ & 99.60 $\uparrow$ & 66.93 $\uparrow$ & 98.10 $\downarrow$ & 89.41 $\uparrow$ & 1.06 $\uparrow$
\end{tabular}
\end{table}

The multimodal WFM with LoRA is strong: it improves over FT2 on four of five tasks, with only small declines on HAS and INTD. Gain is most pronounced on RFS (+5\%). LoRA achieves these results with $5\times$ fewer task-specific parameters, enabling full model sharing and, as a result, better multi-task capability.

\section{Conclusion}
\label{sec:conclusion}

In this work, we introduced the first multimodal wireless foundation model that natively processes raw IQ streams and image-like wireless modalities (e.g., spectrograms and CSI). We proposed a ViT-based masked autoencoder for multimodal wireless inputs. Using masked wireless modeling, the model learns general representations that transfer well across tasks and modalities.
We evaluated the model on five downstream tasks spanning both modalities and demonstrated that 
the model exhibits strong, balanced performance. The multimodal WFM is competitive with single-modality WFMs, and in several cases surpasses their performance. LoRA fine-tuning delivers additional gains while preserving full backbone sharing with a small task-specific parameter budget.
%
We expect further improvements from larger and more diverse pretraining data. Overall, the results make a concrete case that multimodal WFMs are viable and useful, broaden the WFM use cases, and advance the vision of AI-native 6G.


\balance
\bibliography{bibliography.bib}
\bibliographystyle{ieeetr}
\end{document}